\documentclass[12pt,preprint]{aastex}

\def\arcsecpoint{$''\!.$}
\def\deg{$^{\rm o}$}

\received{}
\accepted{}

\shorttitle{Reddening in Akn~564}
\shortauthors{Crenshaw et al.}

\begin{document}

\title{Reddening, Emission-Line, and Intrinsic Absorption Properties in the 
Narrow-Line Seyfert 1 Galaxy Akn 564\altaffilmark{1}}

\author{D.M. Crenshaw\altaffilmark{2,3},
S.B. Kraemer\altaffilmark{4},
T.J. Turner\altaffilmark{5,6},
S. Collier\altaffilmark{7},
B.M. Peterson\altaffilmark{7},
W.N. Brandt\altaffilmark{8},
J. Clavel\altaffilmark{9},
I.M. George\altaffilmark{5,6},
K. Horne\altaffilmark{10, 11},
G.A. Kriss\altaffilmark{12},
S. Mathur\altaffilmark{7},
H. Netzer\altaffilmark{13},
R.W. Pogge\altaffilmark{7},
K.A. Pounds\altaffilmark{14},
P. Romano\altaffilmark{7},
O. Shemmer\altaffilmark{13},
and W. Wamsteker\altaffilmark{9},
}

\altaffiltext{1}{Based on observations made with the NASA/ESA Hubble Space 
Telescope. STScI is operated by the Association of Universities for Research 
in 
Astronomy, Inc. under NASA contract NAS5-26555. }

\altaffiltext{2}{Department of Physics and Astronomy, Georgia State 
University, Astronomy Offices, One Park Place South SE, Suite 700,
Atlanta, GA 30303}

\altaffiltext{3}{crenshaw@chara.gsu.edu}

\altaffiltext{4}{Catholic University of America and Laboratory for Astronomy
and Solar Physics, NASA's Goddard Space Flight Center, Code 681,
Greenbelt, MD  20771}

\altaffiltext{5}{Laboratory for High Energy Astrophysics, Code 660, NASA's 
Goddard Space Flight Center, Greenbelt, MD 20771}

\altaffiltext{6}{Joint Center for Astrophysics, Physics Department, University 
of Maryland, Baltimore County, 1000 Hilltop Circle, Baltimore, MD 21250}

\altaffiltext{7}{Department of Astronomy, The Ohio State University, 140 West 
18th Avenue, Columbus, OH 43210}

\altaffiltext{8}{Department of Astronomy \& Astrophysics, The Pennsylvania 
State
University, 525 Davey Laboratory, University Park, PA 16802}

\altaffiltext{9}{ESA, P.O. Box 50727, 28080 Madrid, Spain}

\altaffiltext{10}{School of Physics and Astronomy, University of St. Andrews, 
St. Andrews, KY16 9SS, UK}

\altaffiltext{11}{Department of Astronomy, University of Texas, Austin, TX 
78704}

\altaffiltext{12}{Space Telescope Science Institute, 3700 San Martin Drive, 
Baltimore, MD 21218}

\altaffiltext{13}{School of Physics and Astronomy and the Wise Observatory, 
The 
Raymond and Beverly Sackler Faculty of Exact Sciences, Tel Aviv University, 
Tel 
Aviv 69978, Israel}

\altaffiltext{14}{Department of Physics and Astronomy, University of 
Leicester, 
University Road, Leicester, LE1 7RH, UK}

\begin{abstract}

We use {\it Hubble Space Telescope} UV and optical spectra of the narrow-line 
Seyfert 1 (NLS1) galaxy Akn 564 to investigate its internal reddening and 
properties of its emission-line and intrinsic UV absorption gas.  We find that 
the extinction curve of Akn 564, derived from a comparison of its UV/optical 
continuum to that of an unreddened NLS1, lacks a 2200 \AA\ bump and turns up 
towards the UV at a longer wavelength (4000~\AA) than the standard Galactic, 
LMC, and SMC curves. However, it does not show the extremely steep rise to 
1200 
\AA\ that characterizes the extinction curve of the Seyfert 1 galaxy NGC 3227.
The emission-lines and continuum experience the same amount of reddening, 
indicating the presence of a dust screen that is external to the narrow-line 
region (NLR). Echelle spectra from the Space Telescope Imaging Spectrograph 
show 
intrinsic UV absorption lines due to Ly$\alpha$, N~V, C~IV, Si~IV, and Si~III, 
centered at a radial velocity of $-$190 km s$^{-1}$ (relative to the host 
galaxy). Photoionization models of the UV absorber 
indicate that it has a sufficient column (N$_H$ $=$ 1.6 x 10$^{21}$ cm$^{-2}$) 
and is at a sufficient distance from the nucleus (D $>$ 95 pc) to be the 
source 
of the dust screen. Thus, Akn 564 contains a dusty ``lukewarm absorber'' 
similar 
to that seen in NGC 3227.

\end{abstract}


\newpage

\section{Introduction}

Narrow-line Seyfert 1 (NLS1) galaxies were first introduced as a class of 
active 
galactic nuclei (AGN) by Osterbrock and Pogge (1985). Their optical spectra 
show 
narrow (FWHM $<$ 2000 km s$^{-1}$) permitted and forbidden lines, similar to 
Seyfert 2 galaxies, but their emission-line ratios are more like those of 
Seyfert 1 galaxies. In particular, Osterbrock and Pogge found that while the 
permitted lines are only slightly broader than the forbidden lines, these 
objects show strong Fe~II emission and [O~III] $\lambda$5007/H$\beta$ ratios 
$<$ 
3, indicating the presence of high density gas like that in the broad-line 
region (BLR) of Seyfert 1 galaxies. These authors note that the H$\beta$ 
equivalent widths of NLS1s are smaller than typical values for normal Seyfert 
1s, suggesting that they are not just normal Seyfert 1s seen at a special 
viewing angle. 

Renewed interest in NLS1s occurred when it was discovered that they have 
distinctive X-ray properties; they show a steep soft X-ray excess with a 
photon 
index $\Gamma$ $>$ 3 below 1 -- 2 keV, a steep hard X-ray continuum with 
$\Gamma$ 
$=$ 1.9 to 2.6, and rapid large-amplitude X-ray variability on timescales of 
minutes to 
hours (Leighly 1999, and references therein). In addition, optical studies 
have 
established that NLS1s lie at one end of the Boroson \& Green (1992) 
eigenvector 
1 for low-redshift quasars and Seyfert 1 galaxies in that they show relatively 
strong Fe~II emission and weak [O~III] emission (Boller, Brandt, \& Fink 1996).

Akn 564 (z $=$ 0.02467 and D $=$ 99 Mpc, assuming H$_{0}$ $=$ 75 km s$^{-1}$ 
Mpc$^{-1}$) is one of the brightest NLS1s in the X-ray band (Boller et al. 
1996), and shows intrinsic UV absorption lines in {\it HST} spectra (Crenshaw 
et 
al. 1999). In order to explore the varability characteristics of a NLS1 in 
different wave 
bands, we recently conducted a multiwavelength campaign on Akn 564. The 
optical 
monitoring campaign covered the periods 1998 November -- 
1999 November and 2000 May -- 2001 January, and is described by Shemmer et al. 
(2001). The UV campaign with the {\it Hubble Space Telescope} was carried out 
on 
2000 May 9 -- 2000 July 8 and is described by Collier et al. (2001). 
Concurrent 
{\it RXTE} and {\it ASCA} observations are described in Pounds et al. (2001) 
and 
Turner et al. (2001a), respectively. In this paper, we use the {\it HST} 
observations to explore the internal reddening in Akn 564 and the physical 
conditions in the absorbing gas. 

\section{Observations}

{\it HST} spectra of the nucleus of Akn 564 include those obtained with the 
Faint Object Spectrograph (FOS) and the Space Telescope Imaging Spectrograph 
(STIS). The details of these observations are summarized in Table 1. We 
retrieved the flux-calibrated FOS spectra of Akn 564, as well as those of the 
NLS1 galaxy Mrk 493 for comparison, from the {\it HST} archives (see Crenshaw 
et 
al. [1999] for details on the Mrk 493 observations). We 
combined these spectra in their regions of overlap to produce UV/optical 
spectra 
for both Seyferts in the region 1150 -- 6800 \AA. The FOS UV spectra are 
discussed in Crenshaw et al. (1999); Akn 564 shows strong intrinsic 
absorption lines in L$\alpha$, N~V $\lambda\lambda$1238.8, 1242.8, Si~IV 
$\lambda\lambda$1393.8, 1402.8, and C~IV $\lambda\lambda$1548.2, 1550.8, 
whereas 
Mrk 493 shows no evidence for intrinsic absorption. The large wavelength 
coverage of these spectra is particularly useful for determining the reddening 
curve of Akn 564 (\S 3) and the measured fluxes of a large number of 
UV/optical 
emission lines (\S 4) through a fixed aperture (0\arcsecpoint86 in diameter).

As previously mentioned, the STIS spectra were obtained in an intensive 
monitoring campaign in 2000 May -- July. Collier et al. (2001) give the 
details 
of the monitoring sequence and data reduction. For the current study, we used 
the low-dispersion G140L and G230L spectra averaged over 46 visits to assist 
in 
the identification of emission features and continuum regions in the UV. We 
note 
that the UV continuum fluxes varied mildly during this period (the ratio of 
maximum to minimum flux at 1360 \AA\ was 1.29) and that the FOS UV continuum 
fluxes lie within this range; they are lower than those of the average STIS 
values by a factor of only 0.89 $\pm0.02$. The FOS and STIS emission-line 
fluxes 
are also the same to within 10\%.

As part of the STIS campaign, we observed the nucleus of Akn 564 at high 
spectral resolution (7 km s$^{-1}$ FWHM) with the E140M grating (see Table 1).
Our reduction of the E140M spectra included a procedure to remove the 
background light from each order using a scattered light model devised by 
Lindler (1998). The individual orders in each echelle spectrum were spliced 
together in the regions of overlap, and the four individual echelle spectra 
were 
weighted by exposure time and averaged to produce a final E140M spectrum.
This spectrum was used in Collier et al. (2001) to evaluate the contamination 
of the emission lines by the absorption lines. In this paper, we use the 
echelle 
observations to study the nature of the intrinsic UV absorption in this object 
(\S 5).

\section{The Reddening Curve in Akn 564}

In Figure 1, we show the FOS UV/optical spectra of Akn 564 and Mrk 493 (the 
latter has been scaled and offset as described in the caption). The smooth 
curves are spline fits to continuum regions that are free of obvious 
contamination by absorption or emission lines, including the ``little blue 
bump'' between 2200 and 4200 \AA, which consists of a blend of broad Fe~II 
and Balmer recombination emission (Wills, Netzer, \& Wills 1985). The 
continuum 
spectrum of Akn 564 is clearly much flatter than that of Mrk 493 in the UV. 
This suggests that the spectrum of Akn 564 may experience a 
significant amount of internal reddening. This is supported by the findings of 
Walter \& Fink (1993), who conclude that Akn 564 is internally reddened 
based on the large observed H$\alpha$/H$\beta$ ratio ($=$ 4.4) and the small 
ratio of ultraviolet to X-ray continuum flux compared to most Seyfert 1 
galaxies (normal and NLS1s) in their sample. These measurements indicate 
significant internal reddening of the emission lines and UV/optical continuum, 
since reddening due to our Galaxy is small (see below). We present additional 
evidence for internal reddening, based on the He~II 
$\lambda$1640/$\lambda$4686 ratio, in \S4.3.

To determine the reddening curve for Akn 564, we used the FOS spectrum of Mrk 
493 as a template. We assume that the UV/optical continuum of Mrk 493 is 
1) unreddened, except for a small Galactic reddening of
E(B$-$V) $=$ 0.03 (Schlegel et al. 1998), and 2) represents the intrinsic 
continuum of Akn 564. The first assumption is based on the large rise of the 
Mrk 
493 spectrum in the UV, which is typical of unreddened Seyfert 1 galaxies. In 
particular, the continuum shape is nearly identical to that in the STIS 
spectrum 
of the unreddened Seyfert 1 galaxy NGC~4151 (Crenshaw et al. 2001). In 
addition, 
the emission lines (narrow plus broad) of Mrk 493 are essentially unreddened 
after correction for 
Galactic reddening; He~II $\lambda$1640/$\lambda$4686 $=$ 8.0 $\pm$ 1.9, which 
is consistent with the intrinsic ratio of 7 -- 9 from photoionization models 
of 
the NLR (Kraemer et al. 1994) and BLR (Rees, Netzer, \& Ferland 1989)).
The second assumption is based on the similarity of NLS1 
galaxies in a number of properties, including their emission-line 
ratios (Osterbrock \& Pogge 1985; Goodrich et al. 1989) and X-ray 
characteristics (Boller, Brandt, \& Fink 1996; Leighly 1999).
Obviously, this assumption needs to explored with a larger sample 
of simultaneous UV and optical observations of NLS1s, since there is likely to 
be intrinsic differences in the spectral energy distributions (SEDs) of these 
objects (the possible effects of these differences are discussed later in this 
section). 

Most of the continuum and emission-line reddening of Akn 564 must be 
intrinsic, 
since the Galactic reddening is small. The maps of Galactic dust IR emission 
from Schlegel et al. (1998) give E(B$-$V) $=$ 0.06, whereas a power-law plus 
Galactic extinction fit to the UV spectrum yields E(B$-$V) $=$ 0.03 (S. 
Collier 
2001, private communication) . We adopt the latter value (0.03), since 
correction of the 
observed spectrum for E(B$-$V) $=$ 0.06 places the 2200 \AA\ continuum region 
well above a smooth continuum fit (Figure 1). The Galactic reddening is 
small compared to the internal reddening (see below), so this choice has 
little 
effect on our results.

To determine the reddening curve for Akn 564, we follow the technique used by 
Crenshaw et al. (2001) in a study of the reddening and absorption in NGC~3227. 
The continuum fits in Figure 1 were each corrected for Galactic reddening, 
using 
E(B$-$V) $=$ 0.03 and the standard Galactic curve (Savage \& Mathis 1979). We 
can immediately determine the value of E(B$-$V) for the internal reddening: 
E(B$-$V) $=$ 2.5 [log(X$_{B}$)~$-$~log(X$_{V}$)], where X is the ratio of the 
Mrk 493 continuum fit to that of Akn~564 as a function of wavelength, and is 
evaluated at the effective wavelengths of the B (4400 \AA) and V (5500 \AA) 
filters. From this relation, we find that E(B$-$V) $=$ 0.14 mag for the 
internal 
reddening in Akn 564.

We can also determine the reddening curve at any wavelength, relative to V and 
as per convention normalized to E(B$-$V) (Savage \& Mathis 1979):

\begin{equation}
\frac{E(\lambda - V)}{E(B - V)} \equiv
\frac{A_{\lambda} - A_{V}}{A_{B} - A_{V}} =
\frac{log(X_{\lambda}) - log(X_{V})}{log(X_{B}) - log(X_{V})},
\end{equation}

where A$_{B}$, A$_{V}$, and A$_{\lambda}$ are the extinctions in magnitudes at 
B, V, and an arbitrary $\lambda$.

To determine the absolute extinction at any wavelength, we 
need to calibrate the above reddening curve:
\begin{equation}
\frac{A_{\lambda}}{E(B-V)} = \frac{E(\lambda - V)}{E(B-V)} + R_{V},
\end{equation}

by determining the offset R$_{V}$ $\equiv$ A$_{V}$/E(B$-$V). We make use of 
the 
result that the standard Galactic, LMC, and SMC extinction laws have 
essentially 
the same values at $\lambda$~$>$~4500~\AA\ (Crenshaw et al. 2001, and 
references 
therein), and assume that this is the case for Akn 564.
\footnote{It would be preferable to match the continuum curves at $\lambda$ 
$>$ 
7000 \AA\ (Cardelli et al. 1989), but the FOS wavelength coverage for Akn 564 
does not extend this far.}
Thus, R$_{V}$ is the constant we need to add to E($\lambda$$-$V)/E(B$-$V) to 
match these laws at long wavelengths.
This yields R$_{V}$ $=$3.1, which is also the standard Galactic value (Savage 
\& 
Sembach 1979).

Figure 2 shows the extinction laws for the Galaxy, LMC, and SMC, along with 
the 
extinction law for Akn~564 determined in the fashion described above. 
The extinction curve for Akn~564 is unusual in that it diverges from the 
others 
at 4000 \AA. At shorter wavelengths, 
the curve climbs steadily with no evidence for the sharp curvature seen in 
other 
laws in the far-UV; there is also no evidence for a 2200 \AA\ bump, similar to 
the SMC curve (see \S 8). By comparison, the extinction curve for NGC 3227 
also 
begins to diverge from the other curves at 3500 -- 4000~\AA, but rises even 
more 
sharply in the UV than the SMC curve (Crenshaw et al. 2001). 

The principal uncertainty in the reddening curve of Akn 564 is likely the 
assumption that its intrinsic SED is identical to that of Mrk 493 (which, as
noted earlier in this section, is essentially the same as that of NGC~4151 in 
an average state).  
Unfortunately, {\it HST} observations of other unreddened Seyfert 1 galaxies 
covering the full UV/optical range are not available to explore this issue. 
Therefore, we use the extreme variations of NGC~4151 itself, which 
was monitored extensively with {\it IUE}. Clavel et al. (1990) find that the 
ratio of UV to optical continuum F$_{\lambda}$(1450)/F$_{\lambda}$(5000) 
varies 
by a factor of 3.1, with high ratios corresponding to high UV fluxes. However, 
Kaspi et al. (1996) find that a portion of this effect is due to a substantial 
galaxy contribution in the apertures used for optical observations; 
subtracting 
the stellar flux at 5000 \AA\ leads to a ratio that varies by a factor of 2.1. 
Assuming this extreme variation for this ratio, the reddening value and 
uncertainity at 1450 \AA\ is therefore A$_{\lambda}$/E(B$-$V) = 11.1 $\pm$ 2.5.
With this assumption, the reddening curve for Akn 564 could possibly be as 
shallow as the Galactic curve in the UV or somewhat steeper than the LMC 
curve, 
but not as steep as the curves for the SMC or NGC 3227. On the other hand, the 
extinction determined from the observed He II $\lambda$1640/$\lambda$4686 
ratio 
is nearly identical to that derived for the continuum (\S 4.3) and the error 
in 
this ratio is only $\sim$30\%, suggesting that the uncertainty in the 
reddening 
curve is much smaller than the conservative estimate quoted above.

\section{The Emission Lines}

\subsection{Measurements}

We have measured the emission-line fluxes in the FOS spectrum of Akn 564 to 
determine the reddening from the He II lines and for a future investigation 
into 
the physical conditions in the emission-line regions and the spectral energy 
distribution (SED) of the ionizing continuum flux (P. Romano et al., in 
preparation).
As is typical for Seyfert 1 galaxies, there are strong contributions to the 
spectrum from a high-density region responsible for lines such as Fe~II (the 
classic BLR in normal Seyfert 1s) and a lower density region responsible for 
forbidden lines such as [O~III] (the NLR). Since Akn 564 is a NLS1, these 
contributions are strongly blended together. Although the permitted lines tend 
to show broader 
wings than the forbidden lines, which is typical in these objects (Osterbrock 
\& 
Pogge 1985; Goodrich 1989), they are not sufficiently broad to deconvolve the 
broad and narrow components, particularly at this spectral resolution. We 
therefore measured the total emission from each line.

We measured the emission line fluxes of most of the lines by direct 
integration 
of their fluxes over the continuum fit. For severe blends (e.g., H$\alpha$, 
[N~II] $\lambda$6548, 6584), we used the [O~III] $\lambda$5007 and H$\beta$ 
profiles as templates for the forbidden and permitted lines, respectively, to 
deblend the emission features. For the UV lines affected by absorption, we 
used 
the emission-line fits from the STIS echelle spectra (see \S 5). We determined 
the line ratios relative to H$\beta$ and 
corrected these ratios for reddening using the Akn 564 curve in Figure 2, with 
E(B$-$V) $=$ 0.14 ($\pm$0.04), and the standard Galactic curve, with E(B$-$V) 
$=$ 0.03 (our justification for using the Akn 564 continuum reddening curve is 
given in \S 4.3.) We determined errors in the dereddened ratios from the sum 
in 
quadrature of the errors from photon noise, different reasonable continuum 
placements, and reddening. 

We show the observed and dereddened line ratios in Table 2. 
The C~II $\lambda$1334 emission is undetectable in this object, whereas it is 
clearly seen in the UV spectra of most Seyfert galaxies (Crenshaw et al. 
1999). 
There are no Galactic or intrinsic absorption features that could mask the 
C~II 
emission. [O~III] $\lambda$2321 and C~II] $\lambda$2326 are not evident as 
well, 
but their absence could be explained by the presence of Galactic Fe~II 
$\lambda\lambda$2374, 2382 absorption at the redshifted positions of these 
emission lines.

\subsection{Spatial Extent of the NLR}

To place a limit on the size of the NLR in Akn 564, we examined the 
two-dimensional STIS spectral images (there are no narrow-band {\it HST} 
images). We find that the spatial profiles of the continuum and strong 
emission 
lines (L$\alpha$, Mg~II) along the slit are identical and indistinguishable 
from 
a point source (the spatial resolution is 0\arcsecpoint05 -- 0\arcsecpoint1 
FWHM 
in the UV). From the L$\alpha$ spatial profile, we find that at least 85\% of 
the emission-line flux must come from a region that is $\leq$0\arcsecpoint2 
(95 
pc) from the nucleus (assuming circular symmetry). Thus, the NLR is
compact (at least along this direction), which is often the case for Seyfert 1 
galaxies (Schmitt \& Kinney 1996).

\subsection{Reddening of the Emission Lines}

We can investigate the reddening of the emission lines in a manner independent 
of our continuum reddening analysis, since the ratios of the He~II lines vary 
only slightly with temperature and density (Seaton 1978).
The intrinsic He~II $\lambda$1640/$\lambda$4686 ratio is $\sim$7 for the NLR 
(Kraemer et al. 1994) and $\sim$9 for the BLR (Rees et al. 1989).  From this 
range and
the observed ratio of 2.92 (corrected for Galactic reddening), the 
differential extinction is A$_{\lambda}$(1640) $-$ A$_{\lambda}$(4686) $=$
0.95 -- 1.22 mag, By comparison, the continuum extinction curve yields 
A$_{\lambda}$(1640) $-$ A$_{\lambda}$(4686) $=$ 0.95 mag. These values are in 
good agreement, although it is possible that the emission lines experience a 
slight amount of additional extinction (cf., Kraemer et al. 2000a). The 
similar 
extinctions for continuum and emission lines justifies our use of the 
continuum 
reddening curve in \S 4.1. The simplest explanation is a dust screen that lies 
outside of and covers both the continuum source and emission-line regions 
(including the NLR), similar to the situation for NGC 3227 
(Crenshaw et al. 2001). One might expect that the intrinsic UV absorption 
lines 
could arise from gas associated with the dust; this possibility is explored in 
\S 7.

\section{The Intrinsic UV Absorption}

As noted in \S 2, the FOS spectra of Akn 564 show intrinsic absorption from 
L$\alpha$, N~V, Si~IV, and C~IV; these lines are also seen in the 
low-dispersion STIS spectra (Collier et al. 2001). The STIS echelle spectra 
resolve these lines, as shown in Figure 3, which plots the absorption as a 
function of radial velocity (for the stronger member of the doublets) relative 
to the systemic redshift of z $=$ 0.02467, determined 
from the H~I 21-cm line (de Vaucouleurs, et al. 1991). We have also 
discovered the presence of intrinsic Si~III $\lambda$1206.5 absorption at the 
same approximate radial velocity as the Si~IV absorption. However, there is no 
evidence for the presence of lower ionization lines, such as C~II 
$\lambda$1334.5 or Si~II $\lambda$1260.4. It is clear from 
Figure 3 that the L$\alpha$, N~V, and C~IV absorption lines are completely 
saturated, and there is no hope for separating individual kinematic 
components. 
Si~IV and Si~III appear not to be saturated, and common narrow kinematic 
components can be seen in these lines. Due to the noise and inherent 
difficulty in deblending these components, we treat the absorption as a single 
large kinematic component in our analysis. This treatment is appropriate if 
there are no large variations in the physical conditions as a function of 
radial velocity.

To measure the absorption lines, we fitted cubic splines to the continuum and 
emission lines, using the high signal-to-noise low-dispersion spectra as a 
guide. We then divided the echelle spectrum by the spline fit to normalize the 
absorption profiles. For the heavily saturated lines, we obtained a lower 
limit 
to C$_{los}$, the covering of the continuum plus emission by the absorber in 
the 
line of sight, from the residual flux (I$_r$) in the troughs: C$_{los}$ $=$ 
1.0  
$-$ I$_r$. For Si~IV, we determined the covering factor in the deepest portion 
of the lines with the doublet method of Hamann et al. (1997):
\begin{equation}
C_{los} = \frac{I_1^2 - 2I_1 + 1}{I_2 - 2I_1 + 1},
\end{equation}

where I$_1$ and I$_2$ are the residual fluxes in the cores of the weaker line 
(Si~IV$\lambda$1402.8) and stronger line (Si~IV $\lambda$1393.8) respectively.

The derived covering factors are given in Table 3. The values are very close 
to one, which indicates that all of the emission-line gas in this velocity 
range is occulted by the absorber. Since the narrow-line region (NLR) gas 
occupies this velocity range ($\pm$ 500 km s$^{-1}$, based on the optical 
[O~III] lines), {\it the absorber must occult the NLR as well}. For example, 
based on the strengths of the forbidden [Ne~V] lines (Table 2), we know that 
the NLR gas in this object is highly ionized, with an ionization parameter 
similar to that of the compact NLR in NGC 5548 (Kraemer et al. 1998).
If we use the narrow C IV/[O III] ratio from NGC 5548 to estimate the narrow 
C~IV flux in Akn 564, and assume the NLR is 
not covered, the C IV troughs reach an intensity of $\sim$0.5,
instead of the observed value of zero seen in Figure 3.

The radial velocity centroids of the absorption lines are given in Table 3; 
these values are relative to the systemic redshift. The discrepant values can 
be attributed to the effects of saturation. The total column of gas is 
apparently smaller at more positive radial velocities, as indicated by the Si 
lines (assuming constant ionization parameter across the profiles). However, 
saturation will tend to fill in the red 
portions of the other absorption lines and shift the velocity centroids to 
less 
negative numbers. The best values for velocity centroid and full-width at 
half maximum (FWHM) weighted by column density are threfore obtained from the 
unsaturated Si~IV and Si~III 
lines: v$_r$ $=$ $-$194 $\pm$ 5 km s$^{-1}$, FWHM $=$ 180 $\pm$ 20 km 
s$^{-1}$. 
On the other hand, the total velocity coverage of the absorber is best 
obtained 
from the heavily saturated L$\alpha$ line: $-$420 to $+$180 km s$^{-1}$, 
relative to systemic. We note that the low-dispersion STIS and FOS spectra 
show  
no evidence for changes in radial velocity coverage ($\geq$ 80 km s$^{-1}$) or 
equivalent width ($\geq$ 0.2 \AA) of the absorbers over the four-year interval 
between the observations.

Since the dust screen and UV absorber both cover the NLR, it is important to 
ascertain if they arise from the same region, and if they do, determine the 
physical conditions in this region. For this purpose, we have derived the 
ionic 
column densities (or limits) for the absorption. For the unsaturated lines, we 
determined the optical depth ($\tau$) as a function of radial velocity (v$_r$) 
for a covering factor of one: $\tau$ $=$ ln (1/I$_r$). For the saturated 
lines, 
we assumed the minimum possible covering factor (0.97) to derive an absolute
lower limit to  the optical depth using the equation from Hamann et al. (1997):
\begin{equation}
\tau = ln \left(\frac{C_{los}}{I_r + C_{los} -1}\right) 
\end{equation}

We then obtained the ionic column density (or lower limit) by integrating the 
optical depth across the profile:
\begin{equation}
N = \frac{m_{e}c}{\pi{e^2}f\lambda}~\int \tau(v_{r}) dv_{r},
\end{equation}
(Savage \& Sembach 1991), where $f$ is the oscillator strength and $\lambda$ 
is 
the laboratory wavelength (Morton et al. 1988).
We have also estimated upper limits to the two low ionization lines that we 
would expect to be strongest: C~II $\lambda$1334.5 and 
Si~II $\lambda$1260.4. The measured ionic column densites or limits are given 
in 
Table 3, along with model values discussed in the next section.

\section{Photoionization Models of the Absorption}

\subsection{Model Parameters}

Photoionization models for this study were generated using the
code CLOUDY90 (Ferland et al. 1998). We have modeled the absorber
as a matter-bounded slab of atomic gas, irradiated
by the ionizing continuum radiation emitted by the central source.
As per convention, the model
is parameterized in terms of the ionization parameter,
U, the ratio of the density of photons with energies $\geq$ 13.6 eV
to the number density of hydrogen atoms (n$_{H}$) at 
the illuminated face of the slab. By matching the observed ionic column
densities, we can constrain the total hydrogen column density
(N$_{H}$) and ionization parameter for the absorber. Of course,
the value of N$_{H}$ we obtain depends on our assumed elemental abundances
(see below). The model is deemed successful when the predicted ionic columns 
match those observed to better than a factor of 2. As noted in \S 5, 
the strongest UV absorption lines (Ly$\alpha$, C~IV, and N~V)
are completely saturated, hence we only have lower limits to these
ionic column densities. However, we can constrain the physical conditions in 
the absorber based on these limits, the columns of Si~III and Si~IV, and 
the upper limits for Si~II and C~II.

The X-ray continuum of Akn 564 is characterized by a strong excess below
1 keV, which has been variously modeled as thermal bremsstrahlung,
a blackbody, a broad Gaussian, or as an
absorption edge (Brandt et al. 1994; Turner, George, \& Netzer 1999). There is 
evidence for a 
steep underlying power-law continuum, with a photon 
index $\Gamma$ $\approx$ 2.5 (Vaughan et al. 1999). As noted by Brandt et al. 
(1994), an extension of such a power-law greatly overpredicts the UV 
flux, hence the spectrum must flatten at energies below 0.1 keV. 
We based our derived  model SED on the observed
fluxes at 1150 \AA\ and 0.5 keV, assuming that the continuum consists of
power laws of the form L$_{\nu}$ $\propto$ $\nu^{-\alpha}$, where $\alpha$ $=$ 
1.0 at $h\nu$ $<$ 13.6 eV, $\alpha$ $=$ 1.07 between 13.6 eV and 0.5 keV,
and $\alpha$ $=$ 1.6 at $h\nu$ $>$ 0.5 keV (consistent with the 0.7 -- 3 keV 
fluxes measured by Turner et al. [2001a]). To account for the soft X-ray 
excess, 
we also included
a blackbody of temperature $kT$ $=$ 0.157 keV, similar to that derived for the 
luminous NLS1 Ton S180 (Turner et al. 2001b). The blackbody was scaled to
contribute 1/3 of the flux at 0.5 keV, as suggested by Turner et al. (2001a), 
and is the simplest representation of the SED, given the lack of evidence for 
a 
turn-over (Turner et al. 2001a).
Based on this SED, we estimate the central source
luminosity in ionizing photons is $\approx$ 2.0 x 10$^{54}$ s$^{-1}$.
Note that this is $\sim$ 13 times the luminosity of the Seyfert 1 galaxy
NGC 3227 (Kraemer et al. 2000b), and most likely a conservative
lower limit, since we have assumed that the continuum does not peak 
below 0.1 keV. Figure 4 shows the incident and transmitted continua for the 
models (see \S 6.2).

We have also assumed roughly solar abundances (cf. Grevesse \& Anders 1989) by 
number relative to H, as follows:
He $=$ 0.1, C $=$ 3.4 x 10$^{-4}$, N $=$ 1.2 x 10$^{-4}$, O $=$ 6.8 x 
10$^{-4}$,
Ne $=$ 1.1 x 10$^{-4}$, Mg $=$ 3.3 x 10$^{-5}$, Al $=$ 2.96 x 10$^{-6}$,
Si $=$ 3.1 x 10$^{-5}$, P $=$ 3.73 x 10$^{-7}$, S $=$ 1.5 x 10$^{-5}$, Fe $=$ 
4 
x 10$^{-5}$, and Ni $=$ 1.78 x 10$^{-6}$. 
Given the evidence for a large column of dusty gas and the absence of C~II in 
absorption, we have used a standard depletion of 85\% of the carbon onto 
grains (Draine \& Lee 1984). However, the presence of Si~III in absorption 
argues 
against the presence of silicate grains, and we have therefore assumed no 
depletions of other elements.
\footnote{The presence of carbon grains without silicate grains is an 
empirical 
assumption that provides a good match to the observed ionic column densities 
(\S 
6.2). To our knowledge, this situation has not been seen in other types of 
objects, and may reflect the unusual environment around a Seyfert nucleus.
Interestingly, the C~II $\lambda$1334.5 emission line is not detected and the 
C~IV $\lambda$1550 and C~III] $\lambda$1909 emission lines appear to be 
relatively weak in Akn 564 compared to lines such as 
N~V $\lambda$1240, N III] $\lambda$1750 and Si~III] $\lambda$1890 (Table 2). 
Although this could result from a combination of high densities (Kuraszkiewicz 
et al. 2000) and unusual abundance ratios (Wills et al. 1999; Mathur 2000),
another explanation is that the carbon has been removed from gas phase by 
depletion onto dust grains.}  

\subsection {Model Results} 

We modeled the UV absorber as a single-zone and arrived at the following 
best-fit parameters: U $=$ 0.033 and N$_{H}$ $=$ 1.62 x 10$^{21}$
cm$^{-2}$. The predicted ionic columns are listed in Table 3 and, generally, 
provide a good fit to the observed column densities. The
predictions for Ly$\alpha$, C~IV, and N~V exceed the lower limits, while
those for C~II and Si~II are consistent with their upper limits. 
The predicted column for Si~IV is close to the observed value and, although 
the predicted Si~III column is somewhat high, it is within a factor
of 2 of that observed. In Table 4, we list predicted column
densities for the ions C~III, N~III, and O~VI, which can produce
absorption lines observable in the {\it FUSE} bandpass (910 -- 1185 \AA~),
and O~VII and O~VIII, from which
strong bound-free edges and absorption lines associated
with ``warm'' absorbers arise (Reynolds 1997; George et al. 1998). Based on 
the 
most recent {\it ASCA} spectra 
(Turner et al. 2001a), we determined upper limits for the bound-free
optical depths and column densities of O~VII
($\tau_{OVII}$ $\leq$ 0.054, N$_{OVII}$ $\leq$ 2.2 x 10$^{17}$ cm$^{-2}$) 
and O~VIII ($\tau_{OVIII}$ $\leq$ 0.001, N$_{OVIII}$ $\leq$ 1.1 x 
10$^{16}$ cm$^{-2}$), which are consistent with our model predictions. 

\section{The Connection Between UV Absorption and Reddening}   

The total hydrogen column in our model 
would yield a reddening value of E(B$-$V) $=$ 
0.31 for a Galactic dust-to-gas ratio, from the relation given by 
Shull \& Van Steenberg (1985). Our derived 
internal reddening for this object is E(B$-$V) $=$ 0.14,
which is consistent with the lower than Galactic dust-to-gas ratio in our 
model, due to the absence of silicate grains.
In any case, the UV absorber has enough column to produce 
the observed reddening. Given that the UV absorber 
covers the NLR in Akn 564, the dusty gas must lie further than 95 pc
from the active nucleus (see \S 4.2) and have a density of
n$_{H}$ $\leq$ 10$^{3}$ cm$^{-3}$ based on our model parameters.

There are two additional considerations that
must be addressed. First, although the models
show that the UV absorption is consistent with carbon 
depletion, the spectrum of Akn 564 shows no 
evidence for a 2200 \AA\ extinction feature. However, this feature is
typically absent in reddened AGN (Pittman, Clayton, \& Gordon 1999), and, in 
any 
event, its origin is not well understood. Second, since emission-lines will 
arise in the absorber,
its covering factor is constrained by the observed emission-line fluxes.
For example, the absorber we have modeled at a radial distance of 95 pc would 
require a global covering factor of 0.05 to account for all of the observed 
[Ne~V] $\lambda$3426 and 80\% of [O~III] $\lambda$5007. Since there may
be significant [O~III] $\lambda$5007 emission from the gas in which low
ionization lines such as [N~II] $\lambda$6584 and [O~II] $\lambda$3227
form (see Kraemer et al. 2000a), the covering factor 
of the absorber may be somewhat less than this value.

To summarize, the absorber is of sufficiently high ionization
state to produce saturated N~V and C~IV absorption lines. 
We predict O~VII and O~VIII column densities much lower than
that associated with the ``warm'' absorbers detected in {\it ASCA}
spectra of Seyfert 1 galaxies (Reynolds 1997; George et al 1998).
Hence, we arrive at the same conclusion as that for the 
Seyfert 1 galaxy NGC 3227 (Kraemer et al. 2000b; Crenshaw et al. 2001): the 
dust is embedded in a ``lukewarm absorber'' which lies outside of the NLR. 
 
\section{Discussion}

Our motivations for this study were to investigate the intrinsic UV absorption 
revealed in the FOS archive spectra of Akn~564, and observed at high 
resolution 
with STIS during the recent monitoring campaign, and to evaluate the reddening 
of the UV continuum fluxes for a study of the intrinsic SED (P. Romano et al., 
in preparation). In doing so, we have discovered that the intrinsic absorption 
and 
reddening are linked in this object, in that the absorbing gas has a 
sufficient 
column (N$_H$ $=$ 1.6 x 10$^{21}$ cm$^{-2}$) and distance from the continuum 
source (D $>$ 95 pc) to contain the dust that reddens the continuum source, 
BLR, 
and compact NLR. The properties of the absorbing gas in Akn 564 are very 
similar 
to those in the dusty lukewarm absorber in NGC~3227 (N$_H$ $=$ 2.0 x 10$^{21}$ 
cm$^{-2}$, D $>$ 92 pc; Crenshaw et al. 2001), suggesting a common origin.
One interesting difference, however, is the derived 
reddening curve; in NGC 3227, it is much steeper in the UV, suggesting a 
preponderence of small dust grains (Crenshaw et al. 2001). Simultaneous 
UV/optical spectra of additional reddened Seyfert galaxies, covering a broad 
wavelength range (e.g., 1150 -- 10,000 \AA) at high spatial resolution 
($\sim$0\arcsecpoint1, to minimize contamination by the host galaxy) would be 
helpful in exploring the range of reddening curves in Seyferts.

The ionization state of the gas is relatively high in both NGC~3227 (U $=$ 
0.13) 
and Akn~564 (U $=$ 0.032), but not high enough (given the N$_H$ columns) to 
produce observable O~VII and O~VIII edges detectable by {\it ASCA} (although 
possibly detectable through absorption lines in {\it Chandra} spectra). Thus, 
these absorbers do not qualify as ``dusty warm absorbers'', which have been 
claimed in several AGN, including NGC~3227 (Komossa \& Fink 1997). As noted in 
Kraemer et al. (2000b) and Crenshaw et al. (2001), the dusty lukewarm absorber 
provides a more natural explanation, since it can be 
placed at large distances ($>$ 100 pc) from the nucleus and account for not 
only 
the reddening of the continuum source and BLR, but the observed reddening of 
the 
NLR as well. As in the case of NGC~3227 and Akn~564, the dusty lukewarm 
absorber 
can be identified in reddened sources through the detection of strong UV 
absorption lines near the systemic redshift.

Dusty lukewarm absorbers likely originate from a different location than the 
majority of known intrinsic absorption systems in Seyfert 1 galaxies. The 
latter 
tend to be variable and, in many cases, exhibit high outflow velocities (up to 
2100 km s$^{-1}$, Crenshaw et al. 1999), indicating an origin close to the 
nucleus. On the other hand, the dusty lukewarm absorbers in the two objects we 
have studied are located outside of the majority of the NLR emission. 
Interestingly, the galactic disks of both Seyferts are inclined to our line of 
sight ($i =$ 63\deg\ for NGC~3227, $i =$ 59\deg\ for Akn~564; De Zotti \& 
Gaskell 1985), yielding a line of sight with a larger optical depth that 
intercepts more of the host galaxy's disk compared to face-on ($i  
\approx$ 0) Seyfert galaxies. There is ample morphological and kinematic 
evidence that the extended narrow-line region (ENLR) in a Seyfert galaxy is 
ionized gas in the galactic disk (cf., Unger et al. 1987). Thus, it is likely 
that the dusty lukewarm absorber is a highly ionized component in the ENLR 
that 
we see when looking through the host galaxy's disk.

\acknowledgments

This work was supported by NASA through grant number HST-GO-08265.02-A from 
the 
Space Telescope Science Institute, which is operated by the Association of 
Universities for Research in Astronomy, Inc., under NASA contract NAS5-26555.
DMC and SBK acknowledge support from NASA guaranteed time 
observer funding to the STIS Science Team under NASA grant NAG5-4103.
WNB acknowledges support from NASA LTSA grant NAG5-8107.
This research has made use of the NASA/IPAC Extragalactic Database (NED) which 
is operated by the Jet Propulsion Laboratory, California Institute of 
Technology, under contract with the National Aeronautics and Space 
Administration.

\clearpage

\figcaption[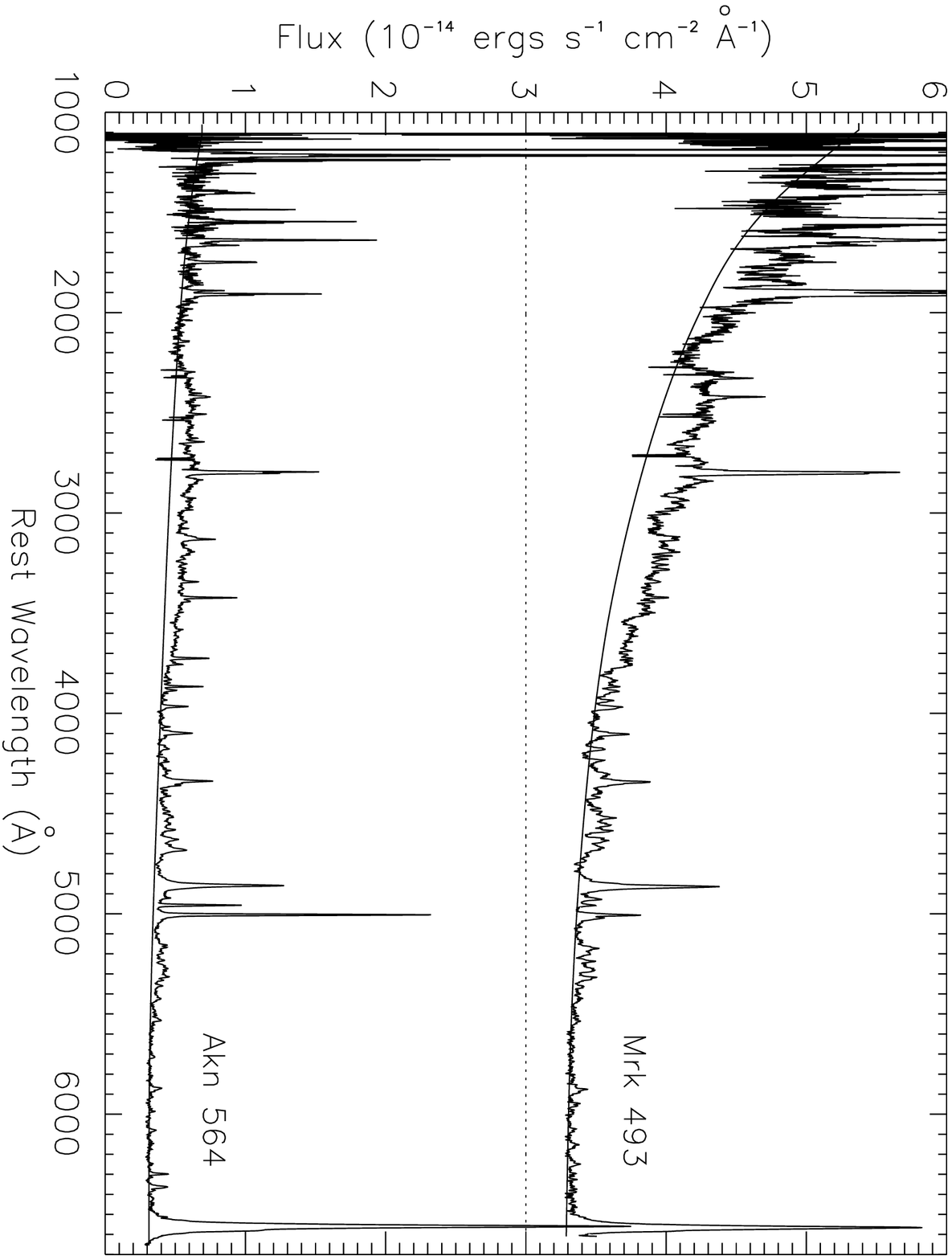]{FOS UV/optical spectra of two narrow-line Seyfert 1 
galaxies. The spectrum of Mrk 493 has been scaled by a factor of 3.0 to match 
the continuum flux of Akn 564 at 5500~\AA, and has been offset by 3.0 x 
10$^{-14}$ ergs s$^{-1}$ cm$^{-2}$ \AA$^{-1}$. The smooth curves are continuum 
fits.}

\figcaption[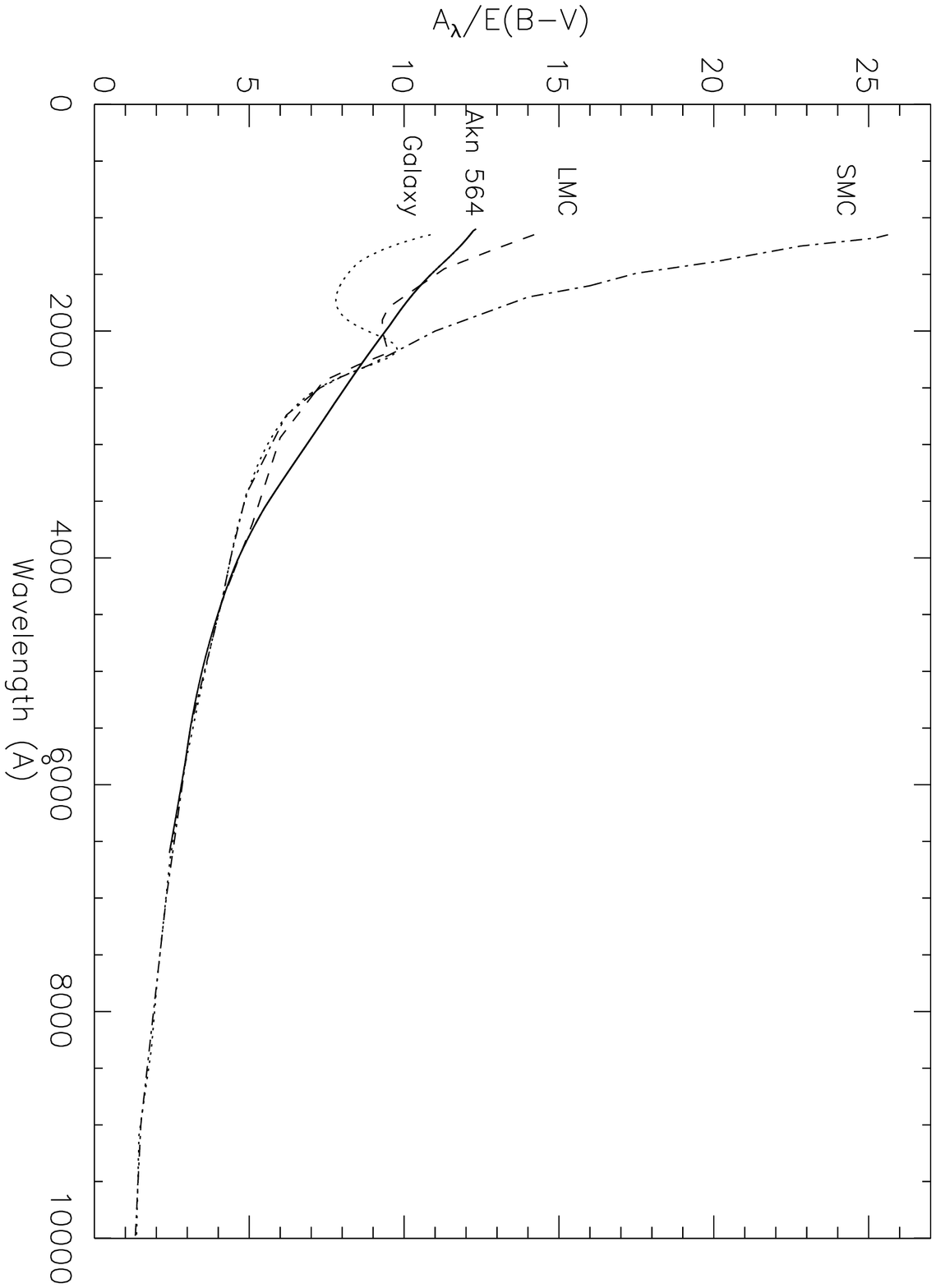]{Reddening curve for Akn~564 as a function of 
wavelength (solid line). Reddening curves for the Galaxy (the standard curve 
of 
Savage \& Mathis 1979), LMC (Koornneef \& Code 1981), and SMC (Hutchings 1982) 
are given for comparison.}

\figcaption[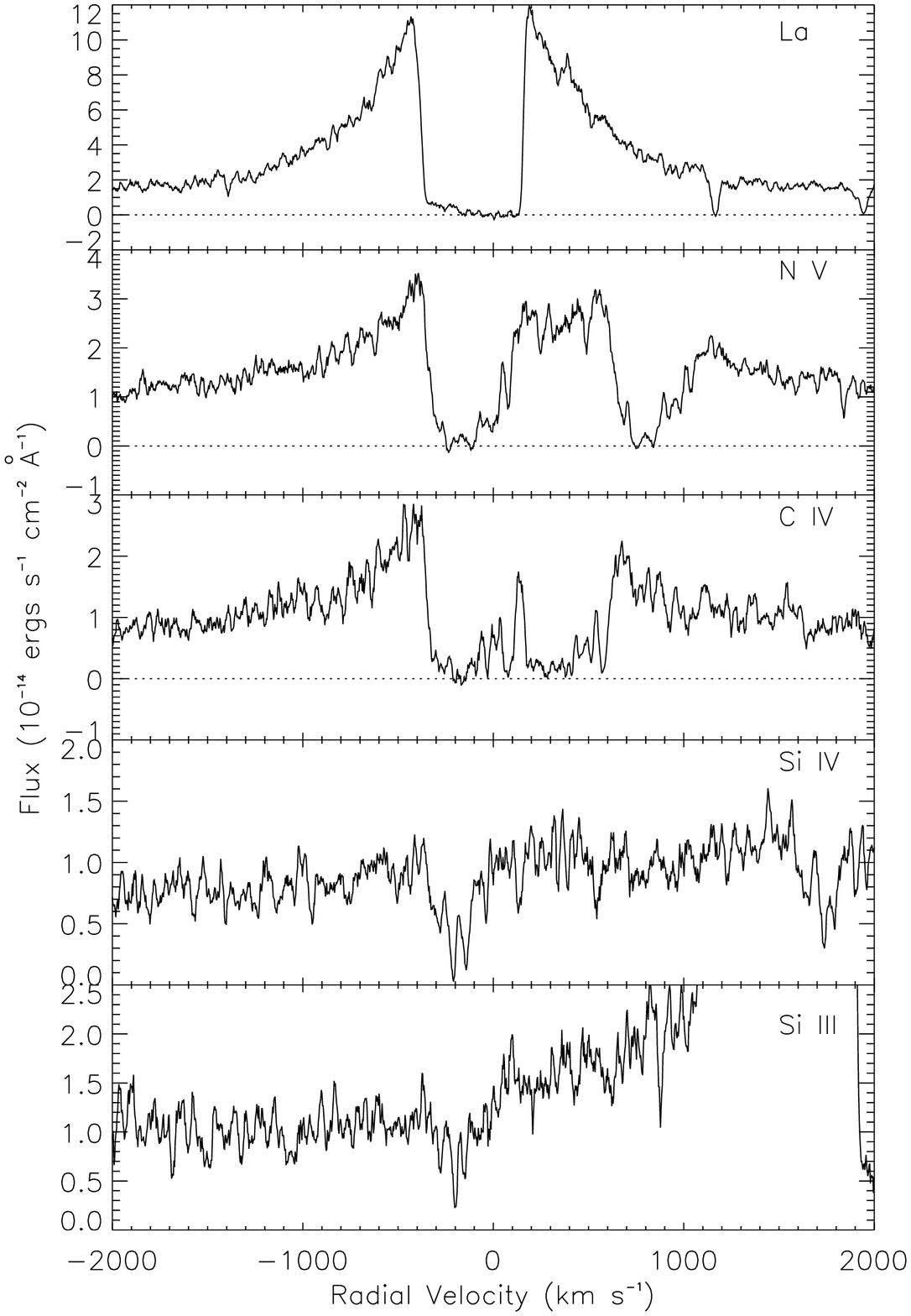]{Portions of the STIS echelle spectra of Akn~564, showing 
the 
intrinsic UV absorption lines detected. Fluxes are plotted as a function of 
radial velocity (of the strongest line for the N~V, C~IV, and Si~IV doublets), 
relative to a systemic redshift of z $=$ 0.02467. The weaker members of the 
doublets at longer wavelengths are also visible.}

\figcaption[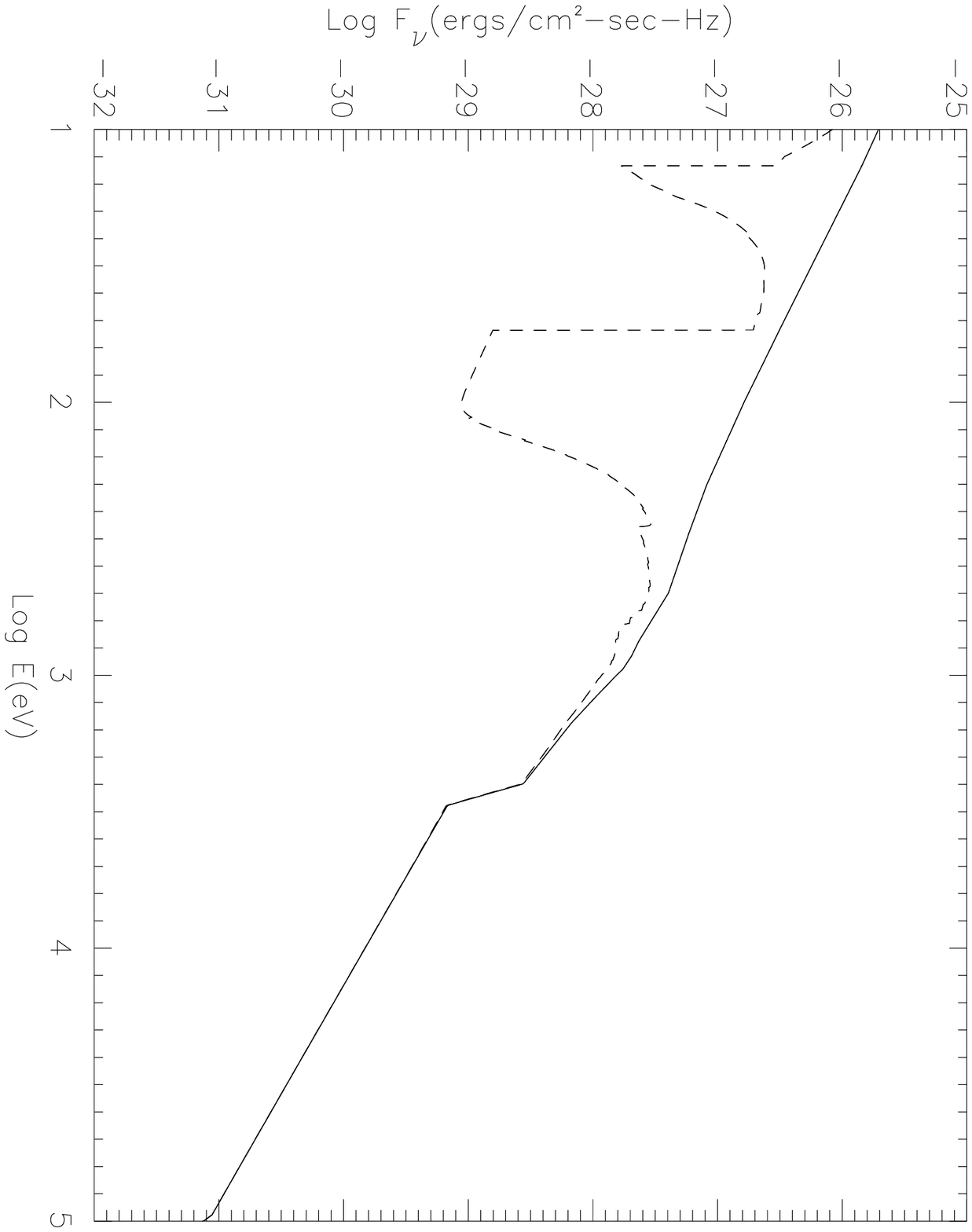]{Incident (solid line) and transmitted (dashed line) 
continua 
for the model described in the text.}

\begin{deluxetable}{ccccrrl}
\tablecolumns{7}
\footnotesize
\tablecaption{{\it HST} Spectra of Akn 564}
\tablewidth{0pt}
\tablehead{
\colhead{Instr.} & \colhead{Grating} & \colhead{Aperture} & \colhead{Coverage} 
&
\colhead{R.P.$^a$} & \colhead{Exposure} & \colhead{Date} \\
\colhead{} & \colhead{} & \colhead{} & \colhead{(\AA)} &
\colhead{($\lambda$/$\Delta\lambda$)} & \colhead{(sec)} & \colhead{(UT)}
}
\startdata

FOS  &G130H &0\arcsecpoint86 diameter &1150 -- 1605 &1300 &3550 &1996 May 23\\
FOS  &G190H &0\arcsecpoint86 diameter &1572 -- 2312 &1300 &1500 &1996 May 23\\
FOS  &G270H &0\arcsecpoint86 diameter &2222 -- 3277 &1300 &1200 &1996 May 23\\
FOS  &G400H &0\arcsecpoint86 diameter &3235 -- 4781 &1300 & 910 &1996 May 23\\
FOS  &G500H &0\arcsecpoint86 diameter &4569 -- 6817 &1300 & 590 &1996 May 23\\
& & & & & \\
STIS &G140L &52$''$ x 0\arcsecpoint5 &1150 -- 1714 &1200 &55,434$^b$ &2000 May 
9 
- July 8\\
STIS &G230L &52$''$ x 0\arcsecpoint5 &1600 -- 3145 &800  &24,216$^b$ &2000 May 
9 
- July 8\\
STIS &E140M &0\arcsecpoint2  x 0\arcsecpoint2 &1150 -- 1710 &46,000     
&10,310 
&2000 May 29 \\
\tablenotetext{a}{Resolving Power ($\lambda$/$\Delta\lambda$) at the central 
wavelength of each spectrum.}
\tablenotetext{b}{Total exposure time for spectra averaged over 46 visits.}
\enddata
\end{deluxetable}

\clearpage
\begin{deluxetable}{lcc}
\tablecolumns{3}
\tablecaption{Akn 564 Emission-Line Ratios
(relative to H$\beta$$^{a}$)}
\tablewidth{0pt}
\tablehead{
\colhead{Line} & \colhead{Observed} & \colhead{Reddening} \\
\colhead{} & \colhead{Ratio} & \colhead{Corrected$^{b}$} 
}
\startdata
L$\alpha$ $\lambda$1216   &4.21 $\pm$0.47 &15.77 $\pm$4.80 \\
N V $\lambda$1240         &1.18 $\pm$0.16 &4.34 $\pm$1.32 \\
O I $\lambda$1302         &0.27 $\pm$0.06 &0.95 $\pm$0.29 \\
O IV] $+$ Si IV $\lambda$1400 &0.35 $\pm$0.07 &1.16 $\pm$0.34 \\
N IV] $\lambda$1486       &0.25 $\pm$0.05 &0.78 $\pm$0.22 \\
C IV $\lambda$1550        &1.13 $\pm$0.16 &3.39 $\pm$0.91 \\
He II $\lambda$1640       &0.55 $\pm$0.08 &1.59 $\pm$0.41 \\
O III] $\lambda$1663      &0.21 $\pm$0.03 &0.60 $\pm$0.15 \\
N III] $\lambda$1750      &0.23 $\pm$0.04 &0.63 $\pm$0.16 \\
Si III] $\lambda$1890     &0.10 $\pm$0.02 &0.25 $\pm$0.06 \\
C III] $\lambda$1909      &0.40 $\pm$0.05 &1.07 $\pm$0.24 \\
$[$Ne IV] $\lambda$2423   &0.09 $\pm$0.02 &0.19 $\pm$0.04 \\
$[$Fe XI] $\lambda$2649   &0.04 $\pm$0.01 &0.07 $\pm$0.02 \\
Mg II $\lambda$2800       &0.58 $\pm$0.08 &1.04 $\pm$0.16 \\
O III $\lambda$3133       &0.18 $\pm$0.06 &0.29 $\pm$0.06 \\
$[$Ne V] $\lambda$3346    &0.06 $\pm$0.01 &0.08 $\pm$0.01 \\ 
$[$Ne V] $\lambda$3424    &0.16 $\pm$0.02 &0.23 $\pm$0.03 \\
$[$Fe VII] $\lambda$3588  &0.02 $\pm$0.01 &0.03 $\pm$0.01 \\ 
$[$O II] $\lambda$3727    &0.12 $\pm$0.02 &0.15 $\pm$0.02 \\
$[$Fe VII] $\lambda$3760  &0.05 $\pm$0.01 &0.06 $\pm$0.01 \\ 
$[$Ne III] $\lambda$3869  &0.11 $\pm$0.01 &0.13 $\pm$0.02 \\
H~8 $+$ He I $\lambda$3889 &0.05 $\pm$0.01 &0.06 $\pm$0.01 \\
$[$Ne III] $+$ H$\epsilon$ $\lambda$3967  &0.11 $\pm$0.02 &0.13 $\pm$0.02 \\
$[$S II] $\lambda$4072    &0.03 $\pm$0.01 &0.03 $\pm$0.01 \\ 
H$\delta$ $\lambda$4100   &0.17 $\pm$0.03 &0.20 $\pm$0.03 \\  
H$\gamma$ $\lambda$4340   &0.32 $\pm$0.05 &0.35 $\pm$0.06 \\
$[$O III] $\lambda$4363   &0.06 $\pm$0.02 &0.06 $\pm$0.02 \\
He II $\lambda$4686       &0.21 $\pm$0.05 &0.22 $\pm$0.05 \\ 
H$\beta$ $\lambda$4861    &1.00           &1.00 \\  
$[$O III] $\lambda$4959   &0.30 $\pm$0.04 &0.30 $\pm$0.04 \\
$[$O III] $\lambda$5007   &0.98 $\pm$0.12 &0.96 $\pm$0.12 \\
$[$Fe VII] $\lambda$5721  &0.04 $\pm$0.01 &0.03 $\pm$0.01 \\
He I $\lambda$5876        &0.08 $\pm$0.02 &0.07 $\pm$0.02 \\
$[$Fe VII] $\lambda$6087  &0.05 $\pm$0.02 &0.04 $\pm$0.02 \\
$[$O I] $\lambda$6300     &0.07 $\pm$0.01 &0.06 $\pm$0.01 \\
$[$Fe X] $\lambda$6374 $+$ [O I] $\lambda$6364 &0.14 $\pm$0.03 &0.12 $\pm$0.03 
\\
H$\alpha$ $\lambda$6563   &4.57 $\pm$0.50 &3.78 $\pm$0.53 \\
$[$N II] $\lambda$6584    &0.15 $\pm$0.05 &0.12 $\pm$0.05 \\

\tablenotetext{a}{Reddening-Corrected Flux (H$\beta$) $=$ 3.13 
($\pm$ 0.34) x 10$^{-13}$ ergs cm$^{-2}$ s$^{-1}$.}
\tablenotetext{b}{Lines corrected using E$_{B-V}$ $=$ 0.14 $\pm$ 0.04 and Akn 
564 reddening curve from Figure~2 plus E$_{B-V}$ $=$ 0.03 and Galactic curve.}

\enddata
\end{deluxetable}

\clearpage
\begin{deluxetable}{lccrr}
\tablecolumns{5}
\footnotesize
\tablecaption{Intrinsic Absorption -- Measurements and Models}
\tablewidth{0pt}
\tablehead{
\colhead{Ion} & \colhead{C$_{los}$} & \colhead{v$_r$$^{a}$}&
\colhead{N (Observed)} & \colhead{N (Model$^b$)}\\
\colhead{} &\colhead{} &\colhead{(km s$^{-1}$)} & \colhead{(cm$^{-2}$)} & 
\colhead{(cm$^{-2}$)} 
}
\startdata
H~I    &1.00 $\pm$0.03 &$-$106 & $>$ 1.4 x 10$^{15}$  & 5.0 x 10$^{17}$\\
N~V    &1.00 $\pm$0.03 &$-$152 & $>$ 3.1 x 10$^{15}$  & 4.7 x 10$^{16}$\\
C~IV   &1.00 $\pm$0.02 &$-$130 & $>$ 2.5 x 10$^{15}$  & 1.4 x 10$^{16}$\\
Si~IV  &0.99 $\pm$0.01 &$-$197 & 1.6 x 10$^{14}$      & 1.5 x 10$^{14}$\\
Si~III &  ------ & $-$190 & 2.6 x 10$^{13}$      & 4.8 x 10$^{13}$\\
C~II   &  ------ & ------ & $<$ 5.4 x 10$^{13}$  & 5.5 x 10$^{13}$\\
Si~II  &  ------ & ------ & $<$ 7.4 x 10$^{12}$  & 1.7 x 10$^{12}$\\
\tablenotetext{a}{velocity centroid relative to the systemic redshift of 
0.02467}
\tablenotetext{b}{U $=$ 0.032, N$_H$ $=$ 1.62 x 10$^{21}$ cm$^{-2}$}
\enddata
\end{deluxetable}

\begin{deluxetable}{lc}
\tablecolumns{2}
\footnotesize
\tablecaption{Additional Predicted Ionic Column Densities}
\tablewidth{0pt}
\tablehead{
\colhead{Ion} &
\colhead{(cm$^{-2}$)} 
}
\startdata
C~III   & 5.2 x 10$^{15}$ \\ 
N~III   & 1.5 x 10$^{16}$ \\
O~VI    & 2.4 x 10$^{17}$ \\
O~VII   & 1.5 x 10$^{17}$ \\
O~VIII  & 5.3 x 10$^{15}$ \\
\enddata
\end{deluxetable}

%\end{document}

\clearpage
\vskip3.0in
\begin{figure}
\plotone{f1.eps}
\\Fig.~1.
\end{figure}

\clearpage
\vskip3.0in
\begin{figure}
\plotone{f2.eps}
\\Fig.~2.
\end{figure}

\clearpage
\vskip3.0in
\begin{figure}
\plotone{f3.eps}
\\Fig.~3.
\end{figure}

\clearpage
\vskip3.0in
\begin{figure}
\plotone{f4.ps}
\\Fig.~4.
\end{figure}

\end{document}